# Chemical basis of deep-Earth physics: Emphasis on the core-mantle boundary D″

J. Marvin Herndon
Transdyne Corporation
11044 Red Rock Drive
San Diego, CA 92131 USA

Communications: mherndon@san.rr.com

**Abstract:** Currently popular ideas about the Earth's interior have developed almost entirely on the basis of physics. In the spirit of the United Nations' designation of 2011 as the International Year of Chemistry, I unify chemical and physical inferences for Earth-matter below the depth of 660 km. I relate by fundamental mass ratio relationships the internal parts of that region with corresponding enstatite chondrite parts, providing a quantitative basis for understanding the "seismically rough" matter at the core mantle boundary, D″, as arising from Earth-core precipitates, the "core-floaters", CaS and MgS. I suggest that the ultra-low velocity zone consists of CaS.

## Introduction

Seismic data, coupled with moment of inertia considerations, can yield information on the existence of physical structures within the Earth and whether they are solid or liquid, but not their chemical compositions, which must come from implications derived from meteorites. Eight years before Oldham [1] discovered the Earth's core from seismic data, Wiechert [2] postulated its existence and its nickel-iron composition from whole-Earth density measurements [3] and from observations of meteorites. Soon after its discovery, the Earth's core was found to be fluid because of its inability to sustain transverse earthquake waves [4].

By the early 1930s, the internal structure of the Earth was thought to be simple, consisting of just the fluid core, surrounded by a uniform shell of solid rock, called the mantle, and topped by a very thin crust [4]. Subsequent seismological evidence, though, indicated more complex structures for both the core and mantle. But the interpretations of these complexities have consistently been made on the basis of physics, generally without regard for chemistry.

Separately the domains of physics and chemistry can provide only partial descriptions of Earth's interior; true understanding must simultaneously and correctly satisfy each domain. In the spirit of the United Nations' designation of 2011 as the International Year of Chemistry, I unify in a self-consistent manner chemical and physical inferences for matter of the deep-interior of Earth with particular emphasis on the region at the core-mantle boundary, commonly referred to as D″, pronounced "dee double prime".

The first indication of Earth-core complexity came with Lehmann's 1936 discovery [5] of the inner core. Birch [6] and others assumed that the inner core was partially-crystallized nickel-iron from the in-process freezing of the nickel-iron fluid core. The explanation proffered was one of changed physical state, not chemical difference.

By 1940, Bullen [7, 8] had recognized a seismic discontinuity in the mantle, an interface where earthquake waves change speed and direction, at a depth of about 660 km, thus separating the mantle into two major parts, upper and lower. Additional seismic discontinuities were later discovered in the upper mantle. Bullen subsequently discovered a zone of seismic "roughness", called D″, located between the core and the seismically-featureless lower mantle [9-12]. Generally, seismic discontinuities within the Earth's mantle, including at D″, have been ascribed to physical changes in a medium of uniform composition, i.e., pressure-induced changes in crystal structure, rather than boundaries between layers having different chemical compositions; physics without chemistry.

**Understanding Chondrite Chemistry**

The fundamental relationships connecting the isotope-compositions of the elements of Earth with those of the chondrite-meteorites, and connecting the abundances of the non-gaseous chemical elements of chondrite-meteorites with corresponding abundances of the elements in the outer portion of the Sun form the basis for knowledge of the chemical and mineral composition of the Earth [13-15]. The similarity of corresponding non-volatile element ratios in the Sun and in chondrites attests to their common origin and to chondrites having had relative simple chemical histories (Figure 1). But not all chondrites are identical; they fall into three distinct groups: *Carbonaceous*, *Enstatite*, and *Ordinary* chondrites. These groups differ primarily in oxygen content, which causes their mineral compositions to be quite different [16-18]. Understanding the origin of those chemical differences reveals processes operant during the formation of the Solar System [19].

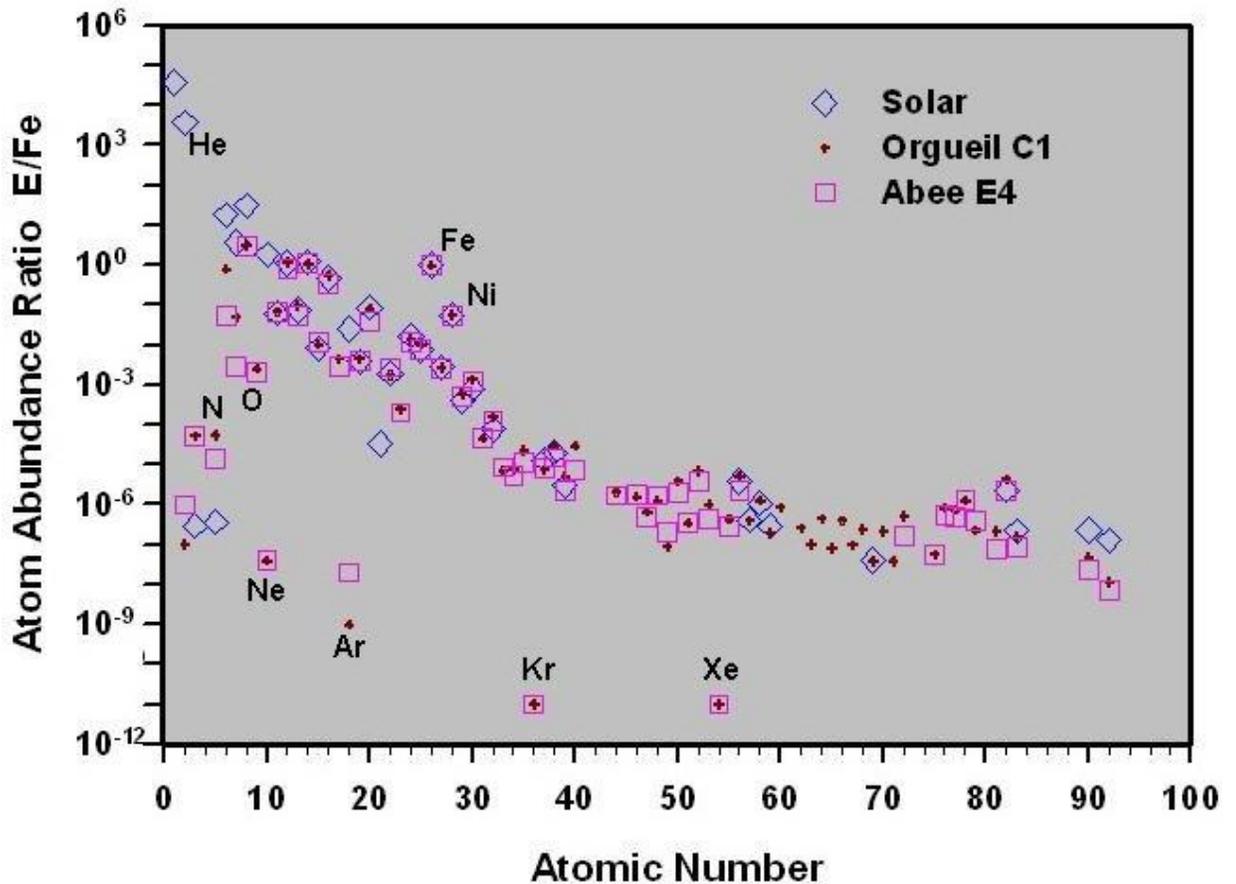

**Figure 1.** Similarity of solar elemental abundance ratios relative to iron with those of the Orgueil carbonaceous chondrite and the highly-reduced Abee enstatite chondrite, demonstrating common element origin and simple chemical histories of the chondrites.

About 90% of the meteorites that are observed falling to Earth are called ordinary chondrites, because they are so common. These are composed of iron metal and silicate-rock, along with some iron sulfide. Many physicists, for more than half a century, have assumed that Earth as a whole resembles an ordinary chondrite [20, 21], which superficially might seem to account for the Earth having an iron alloy core surrounded by a silicate mantle. But ordinary-chondritic-iron-alloy (iron metal plus iron sulfide) is of insufficient relative abundance to account for a core as massive as Earth's. Moreover, the minerals of ordinary chondrites cannot have condensed from an atmosphere of solar composition [22, 23]. Instead, they appear to have formed from two components of matter that previously condensed under different circumstances [24].

The condensation from a gas of solar composition and subsequent separation from that gas phase establishes the oxidation state of the condensate and the planetary matter it becomes.

In a gas of solar composition, the oxidation state of the condensate is primarily governed by the gas phase reaction

$$H_2 + \tfrac{1}{2}O_2 = H_2O$$

which is independent of pressure. Condensation temperature, on the other hand, is a function of pressure. Generally, in solar matter at low pressures, an oxidizable element, for example iron, will condense at low temperatures as an oxide, whereas at high pressures it will condense at high temperatures as a highly reduced condensate [19, 25, 26]. Thus, the oxidation states of the primitive carbonaceous and enstatite chondrites may be understood.

The matter that formed the rare carbonaceous chondrites, like the Orgueil meteorite, appears to have obtained its oxidation state by condensation from a gas of solar composition at low pressures and low temperatures in the outer regions of the Solar System or in interstellar space. In such carbonaceous chondrites, virtually all elements are combined with oxygen. Instead of iron metal, there is magnetite, $Fe_3O_4$. The minerals present, for example, epsomite, $MgSO_4 \cdot 7H_2O$, are indicative of formation at low temperatures. Instead of crystalline silicates, the dominant silicate is an ill-defined, layer-lattice, clay-like mineral. The massive cores of the terrestrial planets preclude their having formed from such oxidized matter. Carbonaceous and ordinary chondrite matter, however, may contribute as an outer veneer-component for Earth, and, especially, for Mars.

Enstatite chondrites, like the Abee meteorite, represent the most highly reduced, naturally-occurring mineral-assemblage known. They contain minerals not found on the surface of Earth, such as oldhamite, CaS. But they do contain copious amounts of iron metal. In the 1940s, during the formative period for deep-Earth physics, the rare enstatite chondrites were not understood and were totally ignored. How their highly reduced state of oxidation came about was a mystery until the mid-1970s.

On the basis of thermodynamic considerations, Suess and I [26] showed that some of the minerals of enstatite meteorites could form at high temperatures in a gas of solar composition at pressures above about 1 bar, provided thermodynamic equilibria are frozen in at near-formation temperatures. At such pressures, molten iron, together with the elements that dissolve in it, is the most refractory condensate. Although there is much to verify and learn about the process of condensation from near the triple point of solar matter, the glimpses we have seen are remarkably similar to the 1944 vision of Eucken [27], i.e., molten iron raining out in the center of a hot, gaseous protoplanet, forming Earth's core, followed by condensation of the silicate mantle.

The Abee enstatite chondrite and others like it have virtually the same relative proportion of volatile trace elements as the Orgueil carbonaceous chondrite (Figure 1). But these enstatite chondrites are replete with high temperature minerals, for example, sub-euhedral crystals of enstatite embayed by iron metal in Abee, and metallurgical evidence of very rapid cooling from a high temperature [28, 29]. Significantly, only enstatite chondrites like Abee have a sufficiently great relative proportion of iron alloy to comprise the interior of a massive-core planet like Earth.

### Deep-Earth's Enstatite Chondritic Composition

Two discoveries made in the 1960s led me to a new quantitative understanding of deep-Earth's chemical composition. These were: (1) The discovery that enstatite chondrite iron metal contains elemental silicon in addition to nickel [30]; and, (2) The discovery of an intermetallic compound, nickel silicide, $Ni_2Si$, mineral-name perryite, in enstatite meteorites [31, 32]. To me the implication was clear. If Earth's core contains silicon, its presence could fully precipitate nickel as nickel silicide, which would then settle downward to form the inner core with precisely the mass observed [33].

In 1979, I progressed through the following logical exercise: If the inner core is in fact the compound nickel silicide, as I had suggested [33], then the Earth's core must be like the alloy portion of an enstatite chondrite. If the Earth's core is in fact like the alloy portion of an enstatite chondrite, then the Earth's core should be surrounded by a silicate shell like the silicate portion of an enstatite chondrite, i.e., $MgSiO_3$, essentially devoid of FeO. But upper mantle silicates contain appreciable FeO. Thus, the enstatite-chondrite-like silicate shell surrounding the core, if it exists, must be bounded by a seismic discontinuity where earthquake waves change speed and direction because of the layers of different chemical compositions. So now here is a prediction that could be tested.

Using the alloy to silicate ratio of the Abee enstatite chondrite [34] and the mass of the Earth's core, by simple ratio proportion I calculated the mass of that enstatite-chondrite-like silicate mantle shell. From tabulated mass distributions [35], I then found that the radius of that predicted seismic boundary lies within about 1.2% of the radius of the major seismic discontinuity at a depth of 660 km which separates the lower mantle from the upper mantle. This logical exercise led me to discover the fundamental quantitative mass ratio relationships connecting the interior parts of the Earth with parts of the Abee enstatite chondrite [36-38], shown in Table 1.

**Table 1. Fundamental mass ratio comparison between the Earth (lower mantle plus core) and the Abee enstatite chondrite.**

| Fundamental Earth Ratio | Earth Ratio Value | Abee Ratio Value |
|---|---|---|
| lower mantle mass to total core mass | 1.49 | 1.43 |
| inner core mass to total core mass | 0.052 | theoretical 0.052 if $Ni_3Si$ 0.057 if $Ni_2Si$ |
| inner core mass to lower mantle + total core mass | 0.021 | 0.021 |
| D″ mass to total core mass | 0.09‡ | 0.11* |
| ULVZ† of D″ CaS mass to total core mass | 0.012⌐ | 0.012* |

\* = avg. of Abee, Indarch, and Adhi-Kot enstatite chondrites
D″ is the "seismically rough" region between the fluid core and lower mantle
† ULVZ is the "Ultra Low Velocity Zone" of D″
‡ calculated assuming average thickness of 200 km
⌐ calculated assuming average thickness of 28 km
data from references [34, 40, 41]

## Chemical Identification of D″

Only five chemical elements, iron, magnesium, silicon, oxygen, and sulfur, account for about 95% of the mass of any chondrite; 98%, if the four minor elements, nickel, calcium, aluminum, and sodium, are included [39]. Figure 2 shows the distribution of those elements between silicate and iron alloy for the Abee enstatite chondrite. Note the occurrence of silicon, calcium and magnesium in the iron alloy portion. These high-oxygen-affinity elements occur in part in the iron alloy because of limited availability of oxygen in these highly reduced meteorites. The oxygen-starved state of oxidation of that enstatite chondrite, and the inner 82% of Earth, was established during condensation from solar matter, where oxygen was more readily bound to hydrogen at high temperatures and high pressures [19].

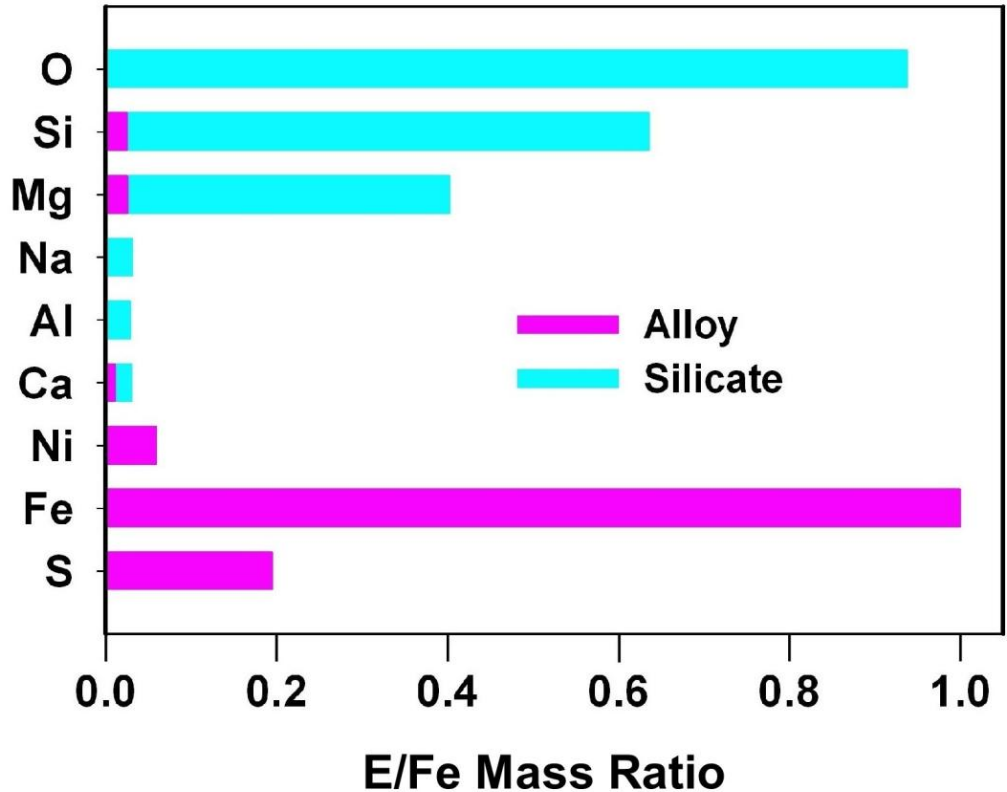

**Figure 2.** Distribution of major and minor elements between silicate and iron-alloy portions of the Abee enstatite chondrite, normalized to iron. Note the high-oxygen-affinity elements calcium, magnesium, and silicon that occur in the iron-alloy portion. These are incompatible elements that precipitate to form the inner core and the matter at the core mantle boundary, D″.

Generally, elements that have a high affinity for oxygen tend to be incompatible in an iron-based alloy. Upon cooling from a high temperature, these incompatible elements will precipitate as soon as thermodynamically feasible. Silicon, I submit, combined with nickel to precipitate as Earth's nickel silicide inner core [33], whereas calcium and magnesium combined with sulfur to form CaS and MgS which floated to the top of the core forming D″ [38, 39]. Note from Table 1 the mass ratio agreement between enstatite chondrite CaS-plus-MgS and Earth's D″ layer.

So, how is it possible to begin to know the behavior of calcium and magnesium in an iron-based alloy? There is an industrial process that is really quite similar. Sulfur impurity can weaken steel. To remove sulfur from high-quality steel, magnesium or calcium is injected into the molten iron to combine with the sulfur at a high temperature, forming either MgS or CaS, which floats to the surface [42, 43]. The same chemical reaction would take place at deep-Earth pressures. Such pressures do not change chemistry, although they might cause a complex molecule to disproportionate into simpler molecules; neither CaS nor MgS is a complex molecule.

Seismic evidence suggests that the matter at D″ is heterogeneously distributed radially and laterally. At ambient pressure, both CaS and MgS are solids at temperatures several hundred degrees above the melting point of pure iron. As solid precipitates from the Earth's fluid core, these "core-floaters" might form non-uniform layers atop the core.

## Ultra-Low Velocity Zone of D″

At the pressures characteristic of the deep-interior of Earth, density at a given temperature is a function almost exclusively of atomic number and atomic mass. Thus MgS is less dense than CaS which is less dense that Earth's iron alloy core. It therefore expected that CaS would layer directly atop the core with MgS layering atop the CaS layer. Together these form D″.

Seismic investigations have revealed a 5-40 km thick layer within D″, adjacent to the Earth's core, called the Ultra-Low Velocity Zone, where earthquake S-wave velocity drops by about 30% and P-wave velocity reduces by about 10% [44]. CaS, I submit, comprises the Ultra-Low Velocity Zone. Qualitatively one might suspect lower earthquake wave velocities in CaS than in MgS because of the heavier atomic weight of calcium. This is an instance where experimental and/or theoretical investigations of wave velocities can be definitive.

In 1862, Nevil Story-Maskelyne [18, 45] discovered, in one of the enstatite-meteorites, a mineral of composition CaS, which he named oldhamite in honor of Thomas Oldham, the first superintendent of the Geological Survey of India. In 1906, Thomas Oldham's son, Richard, discovered the Earth's core [1]. It seems an appropriate tribute that Oldham's core is surrounded by "islands" of oldhamite, named to honor his father.

## Generalizations

Calcium, magnesium, and silicon occur in the alloy of enstatite chondrites and in the Earth's core as a consequence of their highly-reduced oxidation state. For the same reason, uranium occurs in the alloy of the Abee enstatite chondrite [46]. That observation led me to disclose the background, feasibility and evidence of a nuclear fission georeactor at the center of the Earth as the energy source and production mechanism for the geomagnetic field [38, 47, 48].

Revealing the inextricable connection between chemical and physical processes is not only necessary for understanding the Earth [49], but the other bodies of the Solar System as well, their internal compositions and processes, such as magnetic field generation [50]. Chemistry together with physics. That seems quite appropriate for 2011 the International Year of Chemistry.